\documentclass[12pt,a4paper]{article}
\usepackage{graphicx}
\def\as{\alpha_s}

\def\stop{\tilde{t}}
\def\stopbar{\overline{\tilde{t}}}
\def\stoponium{\stop \stopbar}
\def\mstop{m_{\stop}}
\def\ee{e^+e^-}
\def\beq {\begin{equation}}
\def\eeq {\end{equation}}
\def\beqy{\begin{eqnarray}}
\def\eeqy{\end{eqnarray}}

\begin{document}

\begin{titlepage}
\title{{\normalsize{\hfill PACS:13.60.Hb;14.80.-j;14.80.Ly}}\\[25mm]
\textbf{ Estimates of threshold cross section for stoponium 
production at $e^+e^-$ colliders}}
\author{\textbf{N. Fabiano}\footnote{Nicola.Fabiano@pg.infn.it} \\
\textsl{Physics Department of Perugia University and INFN,} \\
\textsl{sezione di Perugia, via A. Pascoli, 06123, Perugia, Italy}}

\date{}
\maketitle
\begin{abstract}
We estimate the total $\ee \to \stoponium$ cross section near threshold
for a Coulombic potential and compare it to the Born approximation. 
The effect of the beam energy spread for present and future $\ee$
colliders is discussed.
\end{abstract}
\end{titlepage}
\section{Introduction}

In the Standard Model it has been shown that bound states can be created 
for every quark but the top 
(see for instance~\cite{GREEN1S,PRR,KUHNTOP,ME,NOI2} and references therein). 
The latter possibility is ruled out due to the high value of the top quark 
mass, which is responsible for its short lifetime. The top quark decays
directly into a $W$ boson and a $b$ quark before being able to create 
a bound state.
In a recent paper~\cite{NOI} it has been  shown that a finite probability
of formation exists for a supersymmetric bound state made out of a squark 
stop and an
antitop squark, for a certain range of $\stop$ mass and some regions of the MSSM
parameter space. In that, the possibility of signal detection at 
an $\ee$ collider with LEP and NLC characteristics has also been investigated 
by means of a Breit--Wigner formula.

A more refined result is needed in the threshold region which is characterised
by low values of the squark velocity $\beta$, i.e.
\beq
\beta = \sqrt{1-\frac{4 \mstop^2}{s}} \ll 1 \qquad .
\label{eq:beta}
\eeq
For this purpose a by now standard Green function approach has been
developed (see~\cite{GREEN1S},~\cite{GREEN1P},~\cite{GREEN21P} and references
therein). 

We will assume that the supersymmetric bound state creation does
not differ from the standard model case, as the relevant interaction is driven
by QCD and controlled by the mass of the constituent squarks~\cite{NOI}. For this
reason the Schr\"odinger Green function technique is suitable for treating
the problem of the scalar bound state. It will be used to compute the bound
state effects on the cross section of stoponium near threshold. We will compare
the results obtained in this manner to the Born cross section estimates for
energies above threshold. The effects of the beam energy spread of the $\ee$
colliders on the computed cross section will be discussed.

\section{The method}
The basic idea of the method is to consider the Schr\"odinger Green function 
equation~\cite{GREEN1S}
\beq
(\mathbf{H}-E) G(\mathbf{x},\mathbf{y},E) = \delta(\mathbf{x}-\mathbf{y}) 
\qquad ,
\label{eq:es}
\eeq
where $\mathbf{H}$ is the Hamiltonian of the system
\beq
\mathbf{H} = - \frac{\nabla_{\mathbf{x}}^2}{2m} +V(\mathbf{x}) \qquad ,
\label{eq:H}
\eeq
and $V(\mathbf{x})$ is the potential for the squarks.
The imaginary part of the derivative of the Green function given 
by~(\ref{eq:es}) taken at the origin is proportional to the cross 
section at threshold~\cite{GREEN1S,GREEN1P,GREEN21P}.
The finite width $\Gamma$ of the state is taken into account by 
the substitution 
\beq
E \to E + i \Gamma \qquad .
\label{eq:nrgy}
\eeq

Unlike the process
\beq
\ee \to t \bar{t} \qquad ,
\label{eq:etotop}
\eeq
the reaction
\beq
\ee \to \stoponium
\label{eq:etostop}
\eeq
proceeds in P--wave, whereas top quarks are produced in S--wave
configuration. This implies the Born level cross section to grow as
$\sigma \sim \beta$ for top production~(\ref{eq:etotop}),
while for process~(\ref{eq:etostop}) of scalars one obtains 
a slower rise, $\sigma \sim \beta^3$.

The threshold cross section of the process $\ee \to  \stoponium$ is given 
by the following  expression~\cite{BEENAKKER,HIKASA}:
\beqy
\sigma(\ee \to \stoponium) = R \; \frac{\pi\alpha^2}{s} & \times &\left[\tilde Q_\gamma^2 + \frac{(v_e^2
+a_e^2)\tilde Q_Z^2}{4\sin^2 2\theta_W}
\frac{s^2}{(s -M_Z^2)^2 + M_Z^2 \Gamma_Z^2} +\right. \nonumber\\
& & \hphantom{\frac{\pi\alpha^2\beta^3}{s}a}\left.+\,\frac{v_e\tilde
Q_\gamma \tilde Q_Z}{\sin2\theta_W} \frac{s (s-M_Z^2)}{(s
-M_Z^2)^2 + M_Z^2 \Gamma_Z^2}\right] 
\label{eq:stopangle}
\eeqy
where $v_e = -1 + 4 \sin^2 \theta_W$, $a_e = -1$, $M_Z$ and $\Gamma_Z$ are the
mass and the total width of the $Z$ boson respectively.
The charges are defined by $\tilde{Q}_{\gamma} = -Q$, 
$\tilde{Q}_Z = (\cos^2 \theta_{\stop} - 2Q \sin^2 \theta_W) \sin 2\theta_W$ , 
with $\theta_W$ being the standard electroweak mixing angle and $\theta_{\stop}$
is the left--right mixing angle.

The $R$ term of equation~(\ref{eq:stopangle}) is 
obtained~\cite{GREEN1S,GREEN1P,GREEN21P} upon taking the imaginary part of the
derivative at the origin of the Green function given by~(\ref{eq:es}) 
\beq
 R = \frac{1536}{81} \frac{\pi}{\mstop^4}
\Im \left [ \left . \frac{\partial}{\partial \mathbf{x}} \frac{\partial}
{\partial 
\mathbf{y}} G_1(\mathbf{x},\mathbf{y},E) \right ] 
\right |_{\mathbf{x}=0,\mathbf{y}=0}
\label{eq:R}
\eeq
($\Im$ stands for the imaginary part, while $G_1$ is the $l=1$ component of
the Green function).

For the purpose of our investigation, we will use a Coulombic potential for the
Hamiltonian given in~(\ref{eq:H})
(here $r =|\mathbf{x}|)$
\beq
V = - \frac{4}{3} \frac{\as}{r}
\label{eq:coul}
\eeq
where $\as$ is given by the QCD two--loop expression~\cite{2LOOPS}

\beq
\alpha_s(Q^2)= \frac{4 \pi}{\beta_0 \log \left [ Q^2/\Lambda^2_{\overline{MS}}
\right ] } \left \{ 1-\frac{2 \beta_1}{\beta_0^2} \frac{\log \left [ \log
\left [ Q^2/\Lambda^2_{\overline{MS}} \right ] \right ]}{\log \left [ 
Q^2/\Lambda^2_{\overline{MS}} \right ] } \right \}
\label{eq:2loops}
\eeq
and is calculated at a fixed value of the Bohr radius $r_B$ given by 
\beq
r_B=\frac{3}{2\mstop\as} \qquad. 
\label{eq:rb}
\eeq
The validity of this choice has been shown (see ~\cite{NOI2,NOI,IO} and 
references therein), and is essentially justified by the fact that the 
stop quark should be much heavier than all other quarks except (maybe) the top.
The high value of $\mstop$ implies from~(\ref{eq:rb}) that the average distance
between two squarks inside the scalar bound state is small, and therefore the
dominant term of the potential is the Coulomb expression given 
in~(\ref{eq:coul}).
This assumption for the bound state potential allows us to obtain an analytic 
expression for the Green function needed for the threshold cross 
section~\cite{PP}.

Following the above cited authors we introduce some standard notations: 
$E= \sqrt{s} -2 \mstop$ is the energy
displacement from  threshold, $k^2 = -\mstop E$, $\lambda = 3 \as \mstop /2$
is the wavelength, and $\nu = \lambda/k$ is the wave number.
Here the argument of $\as$ is taken to be at the soft scale $1/r_B$ . 
The finite width $\Gamma$ of the bound state is taken into account by means 
of the substitution given in~(\ref{eq:nrgy}).

The expression for the $l=1$ Green function for the Coulombic 
potential~(\ref{eq:coul}) is given by

\beqy
G_1(0,0,k) = \qquad  \qquad \qquad \qquad \qquad &  &  \nonumber \\
= \frac{\mstop}{36 \pi} \lambda \left \{ 2(k^2 - \lambda^2) \left [
\frac{k}{2 \lambda}  + \ln \left ( \frac{k}{\mu_f}\right ) + 2 \gamma_E -
\frac{11}{6} + \psi_1(1-\nu)\right ] + \frac{k^2}{2}\right \} & &
\label{eq:green}
\eeqy
where $\gamma_E$ is Euler constant ($\simeq 0.57721$), and $\psi_1$ is the
digamma function, $\psi_1(x) = d(\ln \Gamma(x))/dx$.
The derivative at the origin of~(\ref{eq:green}) is obtained by the simple 
multiplicative relation
\beq
\left .\frac{\partial}{\partial \mathbf{x}} \frac{\partial}{\partial 
\mathbf{y}} 
G_1(\mathbf{x},\mathbf{y},k) \right |_{\mathbf{x}=0,\mathbf{y}=0} = 9G_1(0,0,k)
\qquad .
\label{eq:diffgreen}
\eeq

Some \textit{caveats} (as described in~\cite{PP}) have to be considered
for the case of 
the $P$ wave. There exists a constant linear term in the decay width
$\Gamma$ contributing to the $l=1$ Green function that cannot be
properly computed in a purely nonrelativistic framework~\cite{GREEN1P}. 
The term independent from $k$ that has to be added to the Green 
function~(\ref{eq:green}) is given by
\beq
0.185 \frac{\mstop^3}{36 \pi} \Gamma
\label{eq:greenlin}
\eeq
and from~\cite{PP} the $\mu_f$ argument of the logarithmic term 
in~(\ref{eq:green}) is given by $0.13 \mstop$ by means of an analysis 
of relativistic perturbation theory. We note that the 
last two results have been obtained only
for the top quark case, and we will assume that those results remain valid also
for the stop quark which has a high mass, presumably close to that of the 
top quark. In any case the two aforementioned terms do not contribute 
much to the Green function estimates, since they are dominated by the 
leading $k^3$ term  of~(\ref{eq:green}).

\section{Results and discussion}
Our analysis of the threshold behaviour of the cross section should hold for a
range of mass values and decay widths. For the mass range we will refer to the
current stop mass value limits~\cite{PDG} and the LEP capabilities, while for
the decay widths we have to take into account the formation requirements of
the bound state~\cite{NOI}. 
A criterion for the formation of bound  states is that the creation of 
a hadron can occur only if the level splitting which depends upon the 
strength of the strong  force between the (s)quarks and their relative 
distance~\cite{ME}, is larger than the natural width of the state. 
This means that, if
\beq
\Delta E_{2P-1P} \ge \Gamma
\label{eq:critde}
\eeq
where $\Delta E_{2P-1P}=E_{2P}-E_{1P}$ and $\Gamma$ is the width of the
would--be bound state, then the bound state exists. 
Performing the analysis in this way allows us also to avoid dealing directly 
with  several parameters of the MSSM model which are relevant to stop 
quark decay (apart from the stop mass), namely the ratio of the two Higgs 
vacuum expectation values $\tan \beta$, the Higgs--higgsino mass parameter 
$\mu$  and the wino mass $M_2$~\cite{HIKASA}. 
The only constraint we have to impose is that the decay width be lower than 
about 1 GeV,
necessary for the scalar bound state creation~\cite{NOI} regardless of the
values assumed for the parameters mentioned above.

\begin{figure}[hb]
\begin{center}
\includegraphics[angle=0,width=1.0\textwidth]{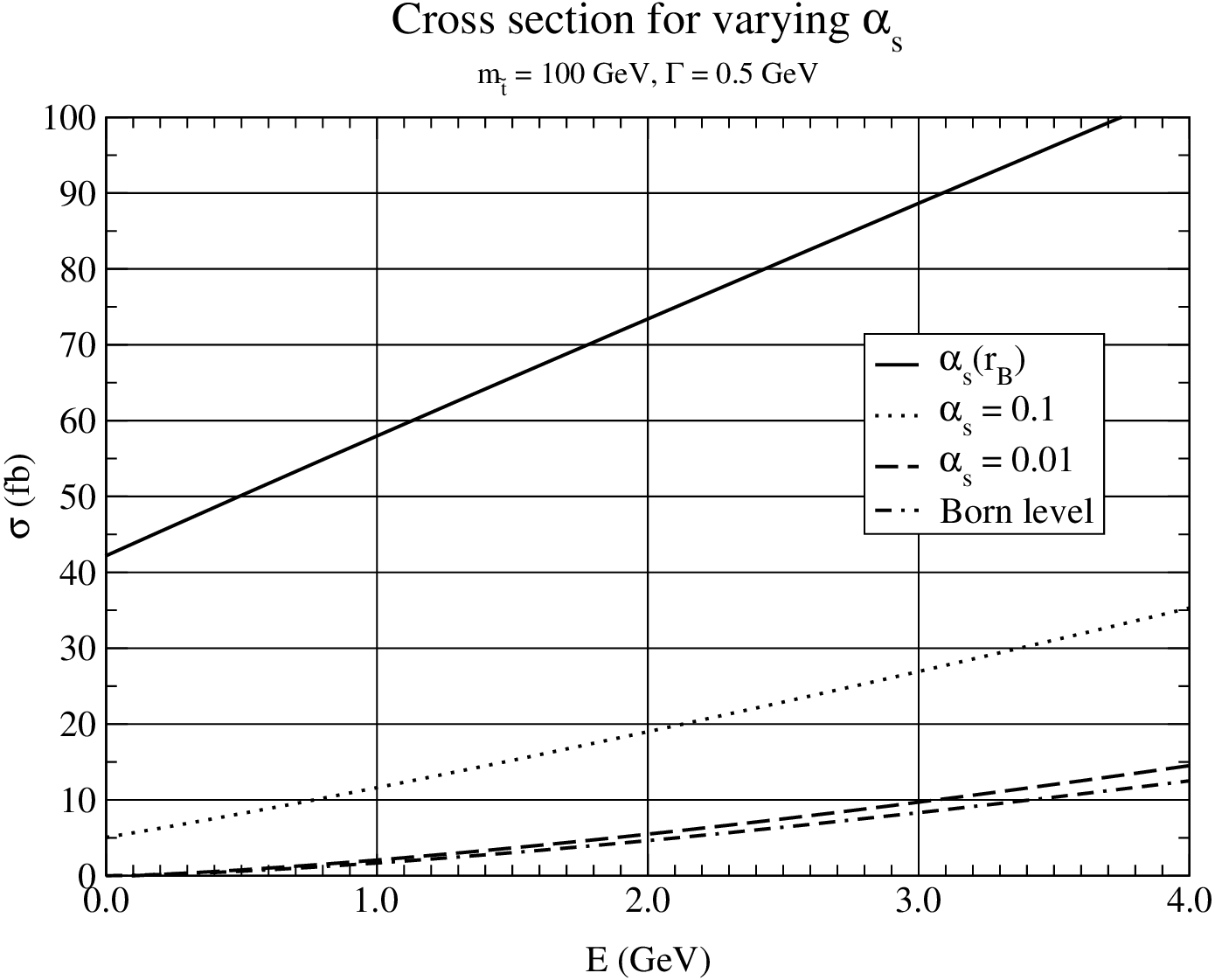}
\end{center}
\caption{ \emph{Comparison of the Green function method and the Born level
expression for the cross section, in the limit $\as \to 0$. Here we assume that
$\cos^2 \theta_{\stop}=1$, $\mstop=100 \;GeV$ and a width of 0.5 GeV.}}
\label{fig:born}
\end{figure}

As a first step, we check that the expression~(\ref{eq:green}) used for our
potential model~(\ref{eq:coul}) is consistent with the Born cross section
for $E>0$. We see in figure~(\ref{fig:born}) that in the non interacting limit 
with $\as \to 0$ the cross section given by~(\ref{eq:green}) tends to the usual
Born expression, and is zero for $E<0$. This confirms the consistency of the
expression found in~(\ref{eq:stopangle}).

\begin{figure}[ht]
\begin{center}
\includegraphics[angle=0,width=0.72\textwidth]{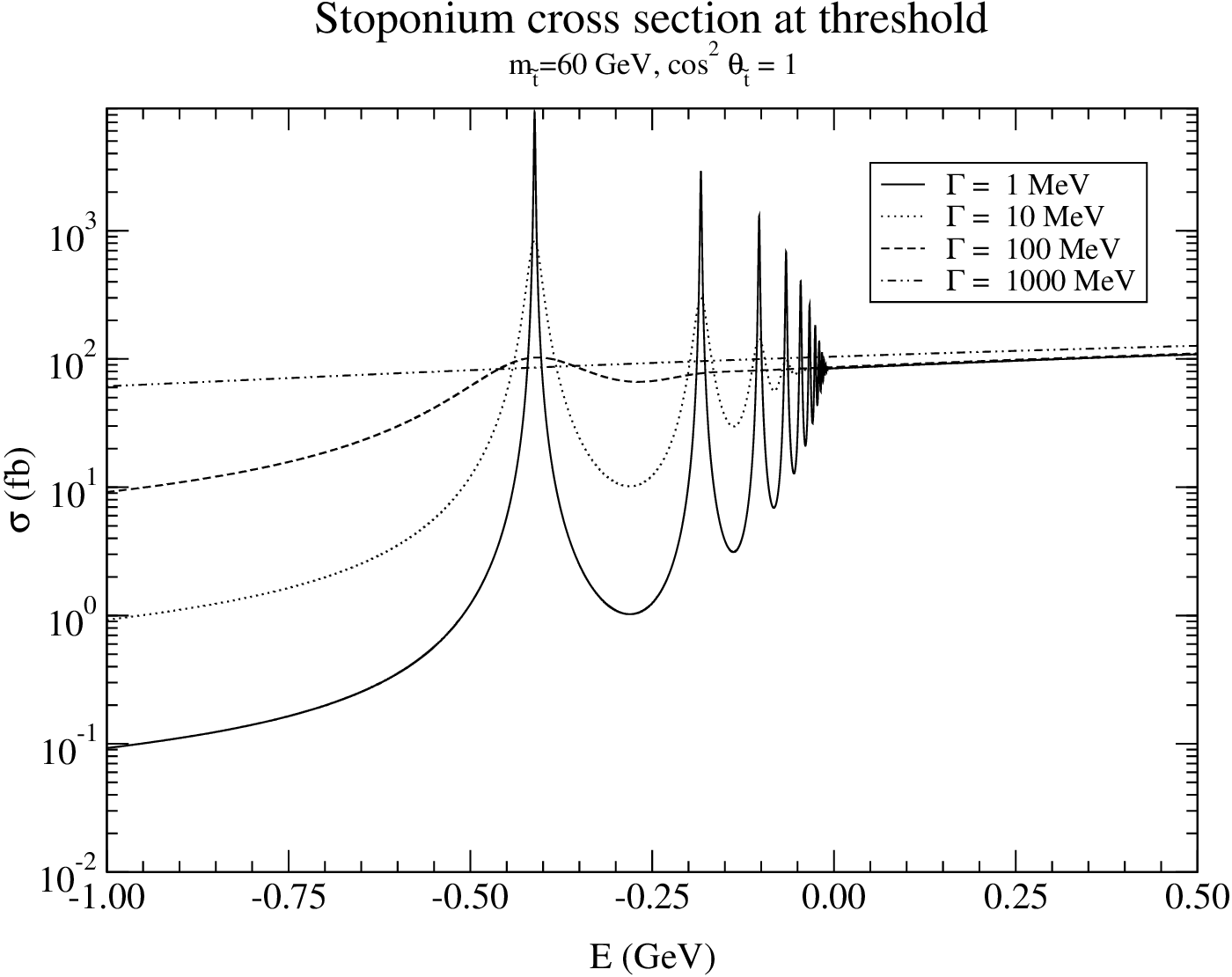}
\end{center}
\caption{ \emph{Cross section at threshold for various decay widths with
$\mstop=60 \; GeV$, $\cos^2 \theta_{\stop} = 1$. The centre of mass energy 
is $\sqrt{s} = 120 \; GeV$ at threshold.}}
\label{fig:Green60}
\end{figure}

\begin{figure}[hb]
\begin{center}
\includegraphics[angle=0,width=0.72\textwidth]{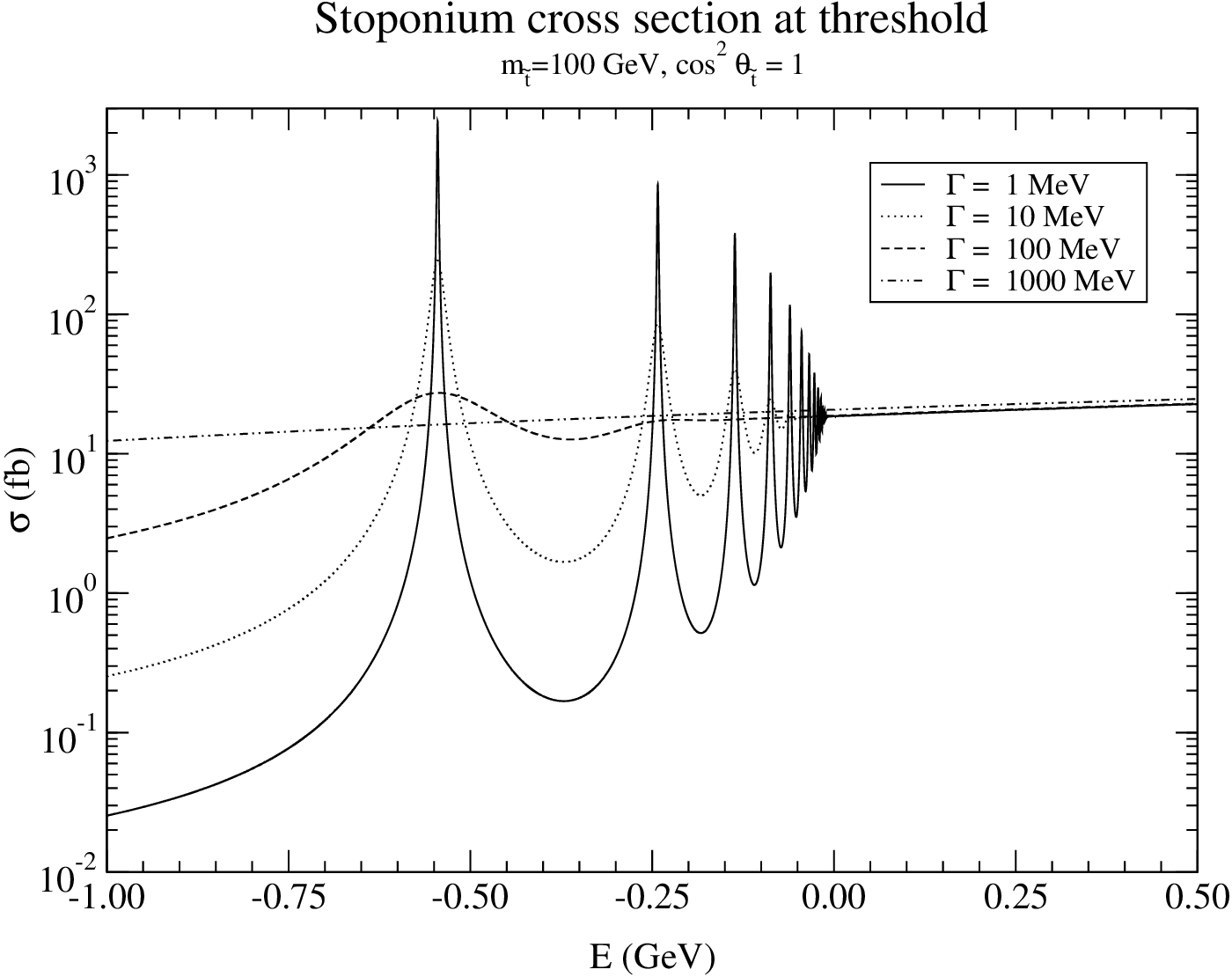}
\end{center}
\caption{ \emph{Cross section at threshold for various decay widths with
$\mstop=100 \; GeV$, $\cos^2 \theta_{\stop} = 1$. The centre of mass energy 
is $\sqrt{s} = 200 \; GeV$ at threshold.}}
\label{fig:Green100}
\end{figure}

In figures~(\ref{fig:Green60},\ref{fig:Green100}) we present the results of the
threshold cross section for a LEP mass range, $\mstop=60, \: 100 \: GeV$. We 
have chosen the left--right mixing angle to be such that 
$\cos^2 \theta_{\stop} = 1$. From~(\ref{eq:stopangle}) it is possible to see 
that the cross section minimum value is obtained for $\mstop=100 \: GeV$ at 
$\cos \theta_{\stop} \simeq \pm 0.55$, the point where the $Z$ boson coupling 
vanishes, which does not much differ from the maximal value.
As  previously stated, assuming the decay width to be smaller
than 1 $GeV$, we show for each chosen mass for widths of $10^{-3},10^{-2},
10^{-1}$ and 1 $GeV$ respectively. The cross sections are plotted against the
threshold offset energy $E$, the centre of mass energy being given by the
relation $\sqrt{s} = 2\mstop +E$.

For $\mstop=60 \; GeV$, figure~(\ref{fig:Green60}) shows the
structure of the discrete energy levels for $E<0$ versus decay width
values given above. The maximal height of the peak, about 8400 $fb$, is 
obtained for the smallest width of $1 \; MeV$. 
The shape of the peaks are similar to the one obtained by the
Breit--Wigner formula, except for the resonance tails that are set higher 
with increasing energy.
The height of the peaks decreases drastically as the binding energy reaches 
asymptotically the
$E=0$ level. They tend also to accumulate and merge towards the $E=0$ value
as the energy increases;
this happens when the distance of the two resonance peaks is of the order of the
decay width. For the Coulombic model the binding energy of the $l=1$ level
is given by the expression
\beq
E_n =  - \frac{4}{9} \frac{\mstop \as^2}{n^2} \quad , \; n>1
\label{eq:energy}
\eeq
and the resonance peaks merge when
\beq
 \frac{4}{9} \mstop \as^2 \left [ \frac{1}{n^2} - 
\frac{1}{(n+1)^2} \right ] \sim \Gamma \qquad .
\label{eq:mpeaks}
\eeq
This means that the the last visible peak has a quantum number $n$ given by
\beq
\frac{2n+1}{n^2 (n^2+1)} \sim \frac{9 \Gamma}{4 \mstop \as^2} \qquad ,
\label{eq:lastpeak}
\eeq
we can see for instance in plot~(\ref{fig:Green60})
that for $\Gamma = 0.1 \; GeV$ the $n=3$ peak is already barely
noticeable. For $\Gamma = 1 \; GeV$, the limiting region of bound state 
formation~\cite{NOI}, we see that there is no visible
structure;  even the first peak is smeared by the large width.

The continuum energy region $E > 0$ does not present any fine
structure, and remains
essentially the same for any given decay width. One important effect 
to be noted is
the large difference of the Green function result with respect to the Born
prediction for $E>0$. This fact is shown in figure~(\ref{fig:born}) where the
two estimates of the cross section are compared. It is clearly seen 
that the former is about one order of
magnitude larger than the Born cross section. This can be understood by the fact
that the Green function technique takes into account the interaction between the
particles, and the contributions of the binding energies accumulate towards 
the $E=0$ energy level as described, thus affecting substantially the continuum 
region as well.

In figure~(\ref{fig:Green100}), we show the results for $\mstop=100 \; GeV$, 
with the same parameters given in figure~(\ref{fig:Green60}). The peaks 
are lower
than the previous case, in particular the largest value of the cross section is
obtained for $\Gamma = 1 \; MeV$ at about 2400 $fb$. The position of the peaks
are shifted towards lower values of the centre of mass energy because 
the binding energy given by the expression~(\ref{eq:energy}) is higher.

\begin{figure}[ht]
\begin{center}
\includegraphics[angle=0,width=0.72\textwidth]{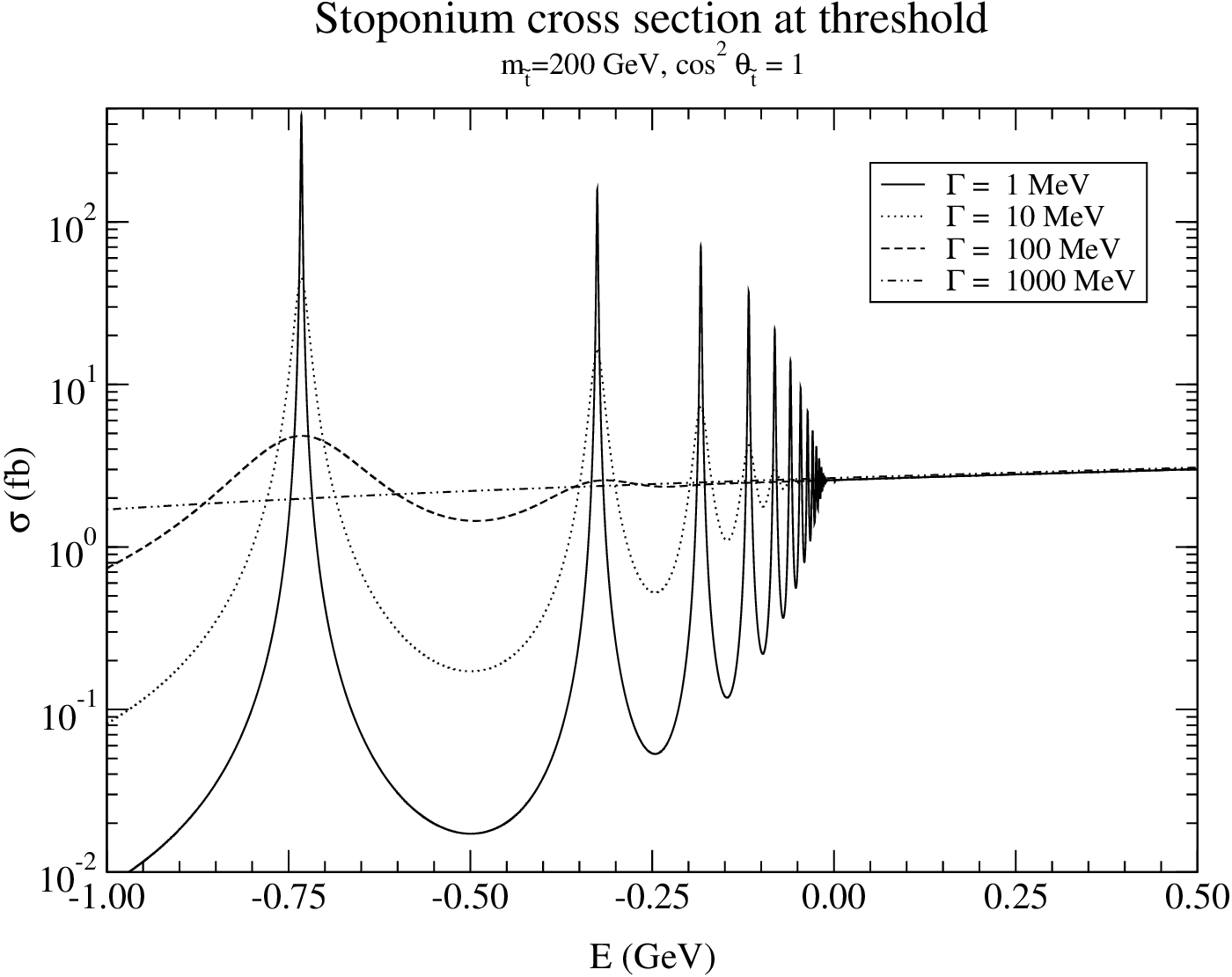}
\end{center}
\caption{ \emph{Cross section at threshold for various decay widths with
$\mstop=200 \; GeV$, $\cos^2 \theta_{\stop} = 1$. The centre of mass energy 
is $\sqrt{s} = 400 \; GeV$ at threshold.}}
\label{fig:Green200}
\end{figure}

\begin{figure}[hb]
\begin{center}
\includegraphics[angle=0,width=0.72\textwidth]{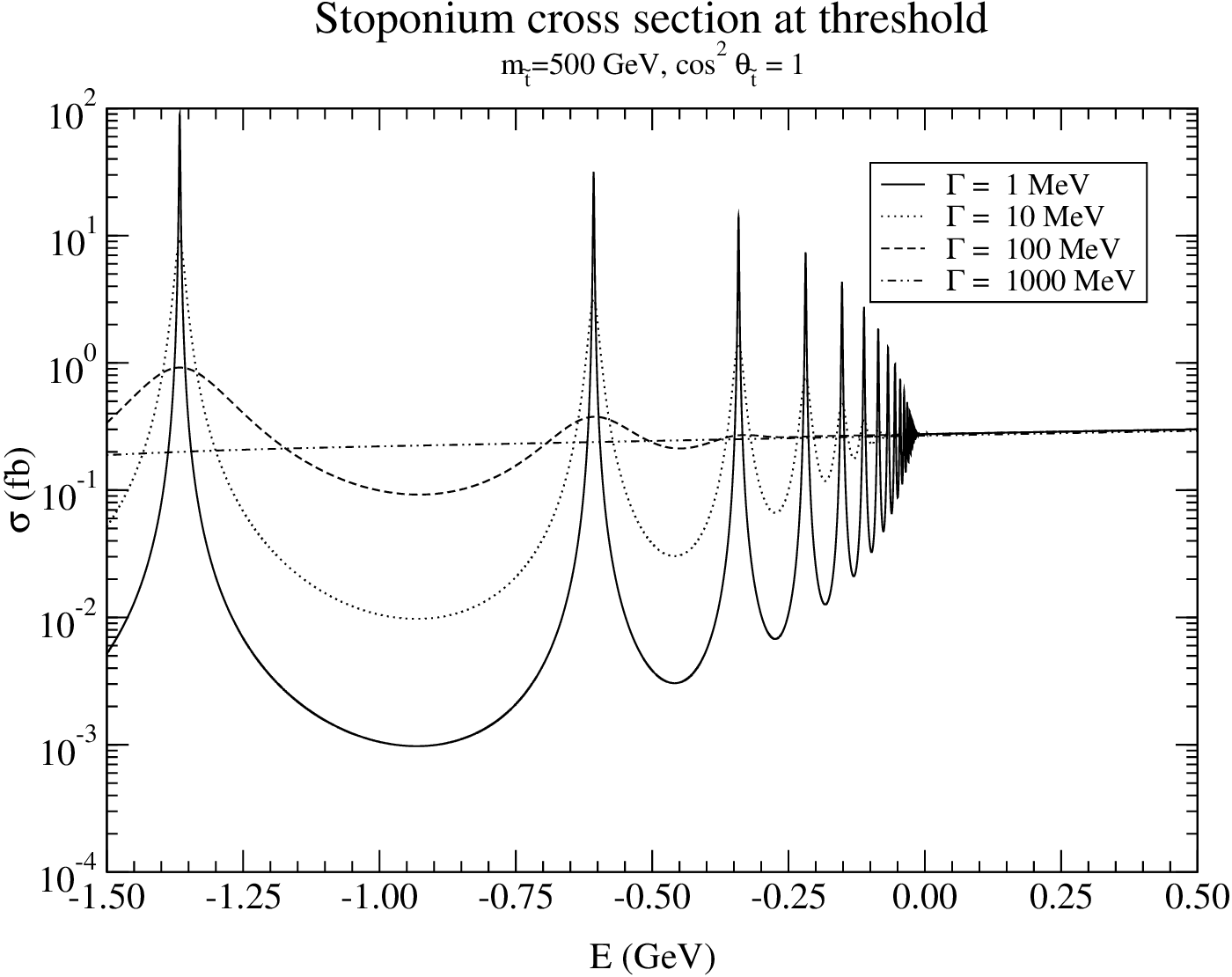}
\end{center}
\caption{ \emph{Cross section at threshold for various decay widths with
$\mstop=500 \; GeV$, $\cos^2 \theta_{\stop} = 1$. The centre of mass energy 
is $\sqrt{s} = 1000 \; GeV$ at threshold.}}
\label{fig:Green500}
\end{figure}

For NLC energies, we present in figures~(\ref{fig:Green200})
and~(\ref{fig:Green500}) results obtained for $\mstop = 200 \; GeV$
and $\mstop = 500 \; GeV$ respectively. The parameters have been chosen to be
the same as of the LEP case, and the qualitative behaviour of the threshold
cross section is analogous. For $\mstop = 200 \; GeV$, the 
maximal value of the first peak obtained at 
$\Gamma =1 \; MeV$ is about 450 $fb$, while for $\mstop = 500 \; GeV$ the
highest peak reaches only 90 $fb$. The position of the peaks is shifted towards
even lower energy values because of still higher values of bound state 
binding energy, as can be verified from~(\ref{eq:energy}).

We remark that all our results have been obtained for the Coulombic
potential~(\ref{eq:coul}). It is known that in the threshold region there are
singular Coulombic terms $(\as/\beta)^n$ which spoil the finite order
perturbation theory. The resummation of these contributions have been done --
see~\cite{PP} and references therein -- and they give small
contributions and only modify slightly the Coulombic
potential. The effect on the cross section is quite small, as can be seen from
the plots in~\cite{PP}, and does not change our estimates by more than a few
percent.

Another point concerns the validity of the Schr\"odinger Green
function method~(\ref{eq:green}). Since it is a 
nonrelativistic procedure, we have to
ensure that the velocity of the squarks is low enough in order to make
relativistic corrections negligible. From kinematical arguments, it is possible
to give a bound for the maximal acceptable  energy offset $E_{MAX}$ . 
Assuming an upper value for the squark velocity, $\beta_{MAX}$ , and
the expression~(\ref{eq:beta}) together with the centre of mass
energy parametrisation, $\sqrt{s}= 2\mstop +E$, one obtains by means of a series
expansion in $E$
\beq
 E_{MAX} < \mstop \beta_{MAX}^2 \qquad .
\label{eq:emax}
\eeq
In table~(\ref{table:en}) we present some estimates on $\beta_{MAX}$ 
and $E_{MAX}$
for different squark masses and some Lorentz $\gamma$ parameter values.
We see that the limit of validity for the nonrelativistic
equation~(\ref{eq:es}) lies in a range of a few $GeV$ around the threshold, 
and naturally increases with larger squark masses; this can be intuitively 
understood since the heavier the squark is, the slower will it rotate inside 
the bound state~\cite{PRR,ME,NOI2,NOI}.

The results obtained so far have to be folded with the beam energy spread
of the collider. For the LEP2 case, which has a beam energy spread of the order
of 200 $MeV$~\cite{PDG}, even if the peak cross section is in the $nb$ range for
$\mstop=60 \; GeV$~(\ref{fig:Green60}) the various resonance peaks are
practically undetectable because their widths are much smaller than the typical
beam energy spread (see also the discussion of~\cite{NOI}). The sole 
possibility of a width larger than the beam
energy spread, $\Gamma = 1 \; GeV$ case, as already discussed, has 
no peaks as they have been smeared and thus no visible fine structure
is envisaged. 
The situation does not change for the value of $\mstop = 100 \; GeV$; as we 
can see from figure~(\ref{fig:Green100}) this situation is essentially the 
same as the previous one, and the peak cross section is even smaller than 
the former by a multiplicative factor of about 4.

With the increase of the centre of mass energy (NLC case) the net result for the
cross section detection is even worse than before. The beam energy spread is of
the order of 6 $GeV$~\cite{TESLA} and, as seen clearly from
figures~(\ref{fig:Green200}) and~(\ref{fig:Green500}), it is even larger than
the energy range used for the plots by a factor of 4, thus making impossible the
detection of any possible fine structure present in the threshold cross section.

\section{Conclusions}

In this letter, we have shown that the bound state effect on the threshold cross
section of a scalar stop bound state is not negligible, at least for the case 
of a Coulombic potential. This effect turns out to be also dramatically
different from the simple Born cross section results for
$E>0$~\cite{HIKASA,DREES}. 

However, because of the large beam energy spread of the present (LEP2) and
future (NLC) $\ee$ colliders, the possible structure of the cross section at
threshold cannot be resolved. This confirmes our less refined
Breit--Wigner approach of~\cite{NOI}, and reinforces our previous
result that the stoponium cannot be detected at the present and even future 
$\ee$ colliders.

\vspace{10mm}
\noindent
{\large \textbf{Acknowledgements}} \\[2mm]

The author wishes to thank M.~Antonelli for useful discussion and support
during this work; and also O.~Panella and P.~Gensini for helpful hints.
Many thanks also to Y.~Srivastava for careful reading and patiently 
correcting  the manuscript.

\newpage

\begin{table}[pt]
\begin{center}
\textbf{{\large Limits of nonrelativistic approach}} 

\begin{tabular}{cccccc}
$\gamma$ & $\beta_{MAX}$ & $E_{MAX} (60)$ & $E_{MAX} (100)$ & $E_{MAX}(200)$ & 
$E_{MAX}(500)$ \\
\hline
1.01 & 0.140 & 1.18 & 1.97 & 3.94 & 9.85 \\
1.02 & 0.197 & 2.33 & 3.88 & 7.77 & 19.42 \\
1.03 & 0.240 & 3.44 & 5.74 & 11.48 & 28.70 \\
\hline
\end{tabular}
\end{center}
\caption{\textit{Some estimates on $\beta_{MAX}$ and $E_{MAX}$ as a function of
the   $\gamma$ parameter for various stop masses, indicated in brackets, in
$GeV$ units.}}
\label{table:en}
\end{table}

\begin{thebibliography}{99}

\bibitem{GREEN1S}
V.S. Fadin, V.A. Khoze, JETP Lett \textbf{46}  525 (1987); \newline
V.S. Fadin, V.A. Khoze, Yad. Fiz. \textbf{48}  487 (1988);\\

\bibitem{PRR}
G. Pancheri, J.P. Revol and C. Rubbia, Phys. Lett. B \textbf{277} 518 (1992).

\bibitem{KUHNTOP}
J.H. K\"uhn, E. Mirkes, Phys. Lett. B \textbf{296} 425 (1992); \newline
J.H. K\"uhn, E. Mirkes, Phys. Rev. D \textbf{48} 179 (1993).


\bibitem{ME}
N. Fabiano, Eur. Phys. J. C \textbf{2} 345 (1998).


\bibitem{NOI2}
N. Fabiano, A. Grau and G. Pancheri, Phys. Rev. D \textbf{50} 3173 (1994); \\
Nuovo Cimento A, Vol. \textbf{107}, 2789, (1994).

\bibitem{NOI}
M. Antonelli, N. Fabiano, \textit{Eur. Phys. J.} \textbf{C16} 361 (2000).


\bibitem{GREEN1P}
V.S. Fadin, V.A. Khoze, Yad. Fiz. \textbf{93} 1118 (1991);\\
I.I. Bigi, V.S. Fadin  and V.A. Khoze, Nucl. Phys. B \textbf{377} 461 (1992);\\
I.I. Bigi, F. Gabbiani and V.A. Khoze, Nucl. Phys. B \textbf{406} 3 (1993);\\
J.H. K\"uhn, E. Mirkes and J. Steegborn, Z. Phys. C \textbf{57} 615 (1993); \\
W. M\"odritsch, Nucl. Phys. B \textbf{475} 507 (1996).
  
\bibitem{GREEN21P}
W. Kwong, Phys. Rev. D \textbf{43} 1488 (1991);\\
M.J. Strassler, M.E. Peskin,  Phys. Rev. D \textbf{43} 1500 (1991);\\
M. Jezabek,  J.H. K{\"u}hn and T. Teubner, Z. Phys. C \textbf{56} 653 (1992);\\
Y. Sumino et al.,  Phys. Rev. D \textbf{47}  56 (1993).

\bibitem{BEENAKKER}
W. Beenakker, R. Hopker and P.M. Zerwas, Phys. Lett. 
B \textbf{349} 463 (1995).

\bibitem{HIKASA}
K. Hikasa, M. Kobayashi, Phys. Rev. D \textbf{36} 724 (1987).


\bibitem{2LOOPS}
W.A. Bardeen, A.J. Buras, D.W. Duke and T. Muta, Phys. Rev.
D \textbf{18}  3998 (1978); \\
W.J. Marciano, Phys. Rev. D \textbf{29}  580 (1984). 

\bibitem{IO}
N. Fabiano, Eur. Phys. J. C \textbf{2} 345 (1998).


\bibitem{PP}
A.A. Penin, A.A. Pivovarov, Nucl. Phys. B \textbf{550}  375 (1999); \\
hep-ph/9904278.


\bibitem{PDG}
Review of Particle Properties, EuroPhys. Journ. C \textbf{3} 1 (1998); \\
http://pdg.lbl.gov/

\bibitem{TESLA}
Ed. R. Brinkmann, G. Materlik, J. Rossbach, A. Wagner,
{\it Conceptual Design of a 500 GeV $e^+e^-$ Linear Collider with Integrated
X--Ray Laser Facility}, Vol. 1 (1997); \\
Ron Settles, private communication; \\
Marcello Piccolo, private communication.

\bibitem{DREES} 
M. Drees, K. Hikasa, Phys. Lett. B \textbf{252} 127 (1990).

\end{thebibliography}
\end{document}